\documentclass[showpacs,prd,floatfix,preprintnumbers,twocolumn,eqsecnum,amsfonts,amssymb,amsmath,tightenlines,superscriptaddress]{revtex4}

\usepackage{graphicx}
\usepackage{amsmath}
\usepackage{amsfonts}   
\usepackage{amssymb}  
\usepackage{mathrsfs}
\usepackage{epsfig}
\usepackage{bm}
\usepackage{color}
\usepackage{natbib}
\usepackage{mystyle}
\usepackage{xcolor}

\begin{document}
\newcommand{\average}[1]{\langle{#1}\rangle_{{\cal D}}}
\newcommand{\dd}{{\rm d}}
\newcommand{\etal}{{\it et al.}}
\newcommand{\super}[1]{$^{#1}$}
\newcommand{\JPU}[1]{\textcolor{blue}{#1}}

\title{A synthetic model of the gravitational wave background from evolving binary compact objects}

\author{Irina Dvorkin}
\email{dvorkin@iap.fr}
\affiliation{Institut d'Astrophysique de Paris, Sorbonne Universit\'{e}s, UPMC Univ Paris 6 et CNRS, UMR 7095, 98 bis bd Arago, F-75014 Paris, France}
\affiliation{Institut Lagrange de Paris (ILP), Sorbonne Universit\'{e}s, 98 bis bd Arago, F-75014 Paris, France}
\author{Jean-Philippe Uzan}
\affiliation{Institut d'Astrophysique de Paris, Sorbonne Universit\'{e}s, UPMC Univ Paris 6 et CNRS, UMR 7095, 98 bis bd Arago, F-75014 Paris, France}
\author{Elisabeth Vangioni}
\affiliation{Institut d'Astrophysique de Paris, Sorbonne Universit\'{e}s, UPMC Univ Paris 6 et CNRS, UMR 7095, 98 bis bd Arago, F-75014 Paris, France}
\author{Joseph Silk}
\affiliation{Institut d'Astrophysique de Paris, Sorbonne Universit\'{e}s, UPMC Univ Paris 6 et CNRS, UMR 7095, 98 bis bd Arago, F-75014 Paris, France}
\affiliation{AIM-Paris-Saclay, CEA/DSM/IRFU, CNRS, Univ Paris 7, F-91191, Gif-sur-Yvette, France}
\affiliation{Department of Physics and Astronomy, The Johns Hopkins University, Baltimore, MD 21218, USA}
\affiliation{BIPAC, University of Oxford, 1 Keble Road, Oxford OX1 3RH, UK}


\begin{abstract}
Modeling the stochastic gravitational wave background from various astrophysical sources is a key objective in view of upcoming observations with ground- and space-based gravitational wave observatories such as Advanced LIGO, VIRGO, eLISA and PTA. We develop a synthetic model framework that follows the evolution of single and binary compact objects in an astrophysical context. We describe the formation and merger rates of binaries, the evolution of their orbital parameters with time and the spectrum of emitted gravitational waves at different stages of  binary evolution. Our approach is modular and allows us to test and constrain different ingredients of the model, including stellar evolution, black hole formation scenarios and the properties of binary systems. We use this framework in the context of a particularly well-motivated astrophysical setup to calculate the gravitational wave background from several types of sources, including  inspiraling stellar-mass binary black holes that have not merged during a Hubble time. We find that this signal, albeit weak, has a characteristic shape that can help constrain the properties of binary black holes in a way complementary to observations of the background from merger events. We discuss possible applications of our framework in the context of other gravitational wave sources, such as supermassive black holes.
\end{abstract}
\date{\today}
\maketitle

\section{Introduction}

The recent detection of the gravitational wave (GW) sources GW150914 \citep{2016PhRvL.116f1102A} and GW151226 \citep{2016ApJ...818L..22A} opened the era of gravitational wave astronomy, and has provided the first direct confirmation of the existence of black holes (BHs), and in particular binary BHs (BBH) that merge within the age of the Universe. Based on the rate of BBH mergers inferred from these detections \citep{2016arXiv160203842A,2016arXiv160604856T} many more sources are expected to be discovered in the second and third Advanced LIGO observing runs \citep{2016arXiv160604856T}.

Ground-based interferometers such as Advanced LIGO, which is already gathering data, as well as VIRGO and KAGRA which are expected to become operational in the near future, are sensitive to gravitational waves in the frequency range $\sim 30 - 1000$ Hz, and are designed to detect mergers of BBH and binary neutrons stars (NSs), as well as the gravitational wave background from unresolved mergers of these binary compact objects \citep{2016arXiv160203847T}. Pulsar timing array (PTA) networks \citep{2013CQGra..30v4009K,2015MNRAS.453.2576L} may detect the GW background produced by merging super-massive BHs (SMBH), topological defects such as cosmic strings, and individually resolvable SMBHs in the frequency range $\sim 10^{-9}-10^{-8}$ Hz. The frequency ranges $\sim 10^{-4} - 10^{-1}$ Hz and $\sim 10^{-1} - 10$ Hz will be explored by the space-based eLISA \citep{PhysRevLett.116.231101} and DECIGO \citep{2011CQGra..28i4011K} observatories, respectively, planned to be launched in the next decade. The frequency coverage that will be attained when all of these observatories are operational suggests the possibility of multi-wavelength GW astronomy \citep{2016PhRvL.116w1102S,2016arXiv160501341N}, where the same source can be observed by different observatories as the merger proceeds.

Detections of individual sources, such as GW150914 and GW151226 are invaluable in studying the properties of compact objects and constraining gravity under extreme conditions. The masses and spins of observed BBH already begin to inform astrophysical models of BH formation \citep{2016ApJ...818L..22A,2016arXiv160204531B,2016arXiv160604889A} and future detections may provide information on the equation of state of NSs \citep{2015PhRvD..92b3012A}. Moreover, the waveforms of individual merger events allow to place stringent constraints on extensions to General Relativity \citep{2016arXiv160604856T}. 

Another component that can be detected with GW observatories is the gravitational wave background from unresolved, merging and inspiraling sources. This component will allow to study the compact object population from a different viewpoint, in particular by constraining the distribution of the binary parameters and their formation mechanisms \citep{2016arXiv160203847T}.

The background from unresolved binary compact objects has in general three components: (a) the signal emitted by core-collapse supernovae (SNe) \citep{1985PhRvL..55..891S,1999MNRAS.303..247F}; (b) the contribution from objects that are about to merge (usually referring to inspiral, merger and ringdown phases) \citep{2008PhRvD..77j4017A,2011PhRvD..84h4037A} and (c) the contribution from inspiraling binaries which do not merge during a Hubble time but which still emit gravitational radiation, resulting in a circularisation and shrinking of their orbit. Component (b) is perhaps the most extensively studied, both in the domain of stellar-mass BBH and NSs as well as SMBH in view of its importance for predicting the signal of merger events such as GW150914 and GW151226. While the waveform of a single isolated merge is well understood, many uncertainties remain, in particular regarding the merger rates (which are related to the properties of the progenitors). GW background from stellar-mass BBH is expected to be detected by Advanced LIGO \citep{2016arXiv160203847T}, while the signal from SMBH is beginning to be constrained by PTA experiments and will be further probed by the eLISA satellite \citep{2015MNRAS.453.2576L,PhysRevLett.116.231101}. Finally, the GW signal from SN collapse is difficult to estimate due to uncertainties in the collapse mechanism \citep{2013ApJ...766...43M,2013ApJ...768..115O,2015PhRvD..92f3005C}. 

While the contribution from merging compact binary systems is dominant, most binaries are not expected to merge within a Hubble time. They will, however, emit gravitational radiation while slowly approaching each other and, depending on the merger rate, the source mass and redshift distribution and the initial orbital parameters, might be detectable with future generations of GW observatories.
 
It is important to stress that the evolution of massive stars and compact objects is affected by their environment. Interactions with other stars in a dense star cluster might be an inportant channel for creating heavy stellar-mass BHs \citep{2016arXiv160604889A} and the growth rate of SMBH is clearly related to the properties of its host galaxy (e.g. \citep{2012Sci...337..544V}). 

The complexity of the different astrophysical processes involved in producing the GW background and their vastly different length and time scales lead to great difficulties in constructing a model that can be easily tested against upcoming data. Moreover, it is often challenging to estimate the relative importance of the various uncertainties involved. In this paper we develop a general framework for calculating the GW background from binary compact objects in an astrophysical context. As will be discussed below, many of the ingredients of this calculation are highly uncertain, therefore we tried to construct a modular approach to the problem, allowing to narrow down on one kind of uncertainty at a time. We then apply this approach to inspiraling stellar-mass BBH and binary NSs that have not merged during a Hubble time. 

The core of our method is in describing the number density of binary systems in terms of the continuity equation in the space of orbital parametres of the binary. A similar approach was used by Refs.  \citep{1994MNRAS.268..841B,1995MNRAS.274..115M,2001MNRAS.327..531I} to study the GW background from high-mass binary pulsars in our Galaxy. In this work we go beyond the steady-state solution assumed in these studies and treat multiple source classes.

This paper is structured as follows: section \ref{sec:Model} describes our synthetic approach: we start with some basic definitions in section \ref{sec:general}. We then define the number densities and formation rates of single and binary compact objects and derive the equations for the evolution of binary orbital parameters in section \ref{sec:evo_nums}. We discuss our complete synthetic model in section \ref{sec:complete_set}. Section \ref{sec:GWback} is an application of our approach to the calculation of several GW backgrounds in the context of a particular astrophysical model. In section \ref{sec:spectrum} we review the GW energy spectrum from inspiraling and merging sources, in section \ref{sec:astro} we outline our astrophysical model and in sections \ref{sec:ins_BH} and \ref{sec:ins_NS} we calculate the GW background from inspiraling and merging stellar-mass BBH and inspiraling binary NSs. We conclude in section \ref{sec:discussion}. 


\section{Description of the model}
\label{sec:Model}

\subsection{Gravitational wave background: general definitions}
\label{sec:general}

%


The spectrum of gravity waves is characterized by the dimensionless density parameter \citep{1999PhRvD..59j2001A}
\begin{equation}
\Omega_{\rm gw}(f)=\frac{1}{\rho_c}\frac{\dd\rho_{\rm gw}}{\dd\ln f}
\end{equation}
where $\rho_c=3H_0^2/8\pi G$ is the critical density of the Universe and $f$ is the frequency measured by the observer. It is related to the frequency at emission $f_e$ by
\begin{equation}
f_e =f(1+z)
\end{equation}
where $z$ is the redshift. For a single class of sources, the energy density of the emitted gravitational waves can be expressed as \citep{2011RAA....11..369R}
$$
\Omega_{\rm gw}(f)=\frac{1}{\rho_c c^3}fF(f),
$$
in terms of the integrated flux of energy received by the observer at frequency $f$,
\begin{equation}
 F(f)=\int p(\theta) \frac{\dd {\cal N}(\theta,z)}{\dd z} {\cal F}(f,\theta,z) \dd\theta\dd z
\end{equation}
where $p(\theta)$ is the probability distribution of the parameters of the sources (such as orbital parameters, masses, etc.). The quantity
\begin{equation}
 {\cal F}(f,\theta,z) = \frac{1}{4\pi \chi^2(z)}\frac{\dd E_{\rm GW}}{\dd f_e}(\theta, f_e)
\end{equation}
depends on the GW signal emitted by the source ${\dd E_{\rm GW}}/{\dd f_e}$ and $\chi(z)$ is the comoving radial distance. The number of sources with parameters in the range $[\theta,\theta+\dd\theta]$ per unit time and redshift interval is given by
\begin{equation}
 \frac{\dd {\cal N}(\theta,z)}{\dd z} =\frac{\dd n}{\dd t}(\theta,z)\frac{\dd V}{\dd z}(z)
\end{equation}
where the comoving volume element is:
\begin{equation}
\dd V(z) =\frac{c}{H_0}\frac{\chi^2(z)}{E(z)}\dd\Omega^2\dd z\:,
\end{equation}
the Hubble parameter is $H(z)=H_0E(z)$ and $\dd\Omega^2$ is the unit solid angle.


The above analysis can be generalized to account for multiple types of sources, such as binary BHs, binary NSs etc. so that the total contribution is obtained by summing over all the components $i$ and their respective internal parameters $\theta_i$:
\begin{align}
& \Omega_{\rm gw}(f)= \nonumber\\ & \frac{1}{\rho_c c^3}f \sum_i
\int  \frac{\dd z}{4\pi\chi^2(z)}\frac{\dd V}{\dd z}(z)
\int  \dd\theta_i 
 \frac{\dd n_i}{\dd t}(\theta_i,z)
\frac{\dd E^{(i)}_{\rm GW}}{\dd f_e}(\theta_i, f_e)
\label{eq:Omega_gw}
\end{align}
where the sum is over different types of sources.
The total background thus depends on the following quantities:
\begin{itemize}
 \item the nature of each class of sources $i$ and the relevant set of parameters $\theta_i$ (i.e. masses, spins, binary orbital parameters etc);
 \item the evolution of the comoving number density of each source per unit time, $\frac{\dd n_i}{\dd t}(\theta_i,z)$;
 \item the evolution of the parameters $\theta_i$ with time. We set $\theta_i(z)\equiv\theta_i[\theta_i^{(0)},z,z_0]$. Among the parameters, the metallicity and orbital parameters evolve with time;
 \item the probability distribution function (PDF) of $\theta_i$ at formation, i.e. ${\cal P}(\theta_i^{(0)})$ which determines the density distribution at later times $n_i(\theta_i,z)$. Its form will generally depend on both the relation $\theta_i[\theta_i^{(0)},z,z_0]$ and the time of formation $z_0$ of the source;
 \item the energy spectrum $\dd E^{(i)}_{\rm GW}/\dd f_e$ emitted by source $i$ with parameters $\theta_i$.
\end{itemize}

These ingredients are described in detail below.

\subsection{Evolution of the number density of binaries}
\label{sec:evo_nums}

The goal of our analysis is to provide a general description of the formation and merger rates of binary systems while accounting for the variation in the binary orbital parameters. 

\subsubsection{Formation of the binaries}

We start by modeling the comoving number density of objects of type $X$. Each of these objects is either single or belongs to a binary system. In the latter case its companion can be another object of type $X$ or an object of different type $Y$, where $X$ and $Y$ can be either a BH or a NS. We thus define
\begin{itemize}
 \item $n_X(M,t)$ : the total number density of objects of type $X$ from which the total number density is obtained as
 $$
 \bar n_X(t) = \int n_X(M,t)\dd M.
 $$
 \item $n^{(1)}_X(M, t)$ : the number density of $X$ that are in a single system with mass $M$;
 \item $n^{(2)}_X(M_1,M_2,{\bm w}, t)$ : the number density of $X$ that are in a $XX$ binary system with masses $M_1$ and $M_2$ and orbital parameters ${\bm w}$;
 \item $n^{(1,1)}_{XY}(M_X,M_Y,{\bm w}, t)$ : the number density of $X$ that are in a $XY$ binary system with masses $M_X$ and $M_Y$ and orbital parameters ${\bm w}$;
\end{itemize}
It is clear from these definitions that
\begin{align}
 n_X(M,t) = n^{(1)}_X(M_, t) & + 2\int n^{(2)}_X(M,M_2,{\bm w}, t) \dd M_2\dd^n{\bm w} \nonumber \\ & +  \int n^{(1,1)}_{XY}(M,M_Y,{\bm w}, t) \dd M_Y\dd^n{\bm w}.
\end{align}
In order to describe the evolution of such systems, we need to calculate their rates of formation. We define the following rates:
\begin{itemize}
 \item $R_X(M,t)$ : the total formation rate of objects of type $X$ with masses $M$ at time $t$;
 \item $R^{(1)}_X(M, t)$ : the formation rate of $X$ that are in a single system with mass $M$ at time $t$;
 \item $R^{(2)}_X(M_1,M_2,{\bm w}, t)$ : the formation rate of $X$ that are in a $XX$ binary system with masses $M_1$ and $M_2$ and orbital parameters ${\bm w}$;
 \item $R^{(1,1)}_{XY}(M_X,M_Y,{\bm w}, t)$ : formation rate of $X$ that are in a $XY$ binary system with masses $M_X$ and $M_Y$ and orbital parameters ${\bm w}$;
\end{itemize}
These rates are clearly related to each other and depend on the chosen physical model of stellar evolution.\\

If we assume for the sake of simplicity that both components in a $XX$ binary system always have equal masses then 
$$
n^{(2)}_X(M,M,{\bm w}, t) =n^{(2)}_X(M,M_2,{\bm w}, t) \delta(M-M_2)
$$
so that
\begin{align}
 n_X(M,t) = n^{(1)}_X(M_, t) & + 2 \int n^{(2)}_X(M,M,{\bm w}, t) \dd^n{\bm w} \nonumber \\ & +  \int n^{(1,1)}_{XY}(M,M_Y,{\bm w}, t) \dd M_Y\dd^n{\bm w}.
\end{align}
We further assume that a fraction $\gamma_X$ of the $X$ component resides in $XY$ systems. We define this ratio by
\begin{equation}
 \gamma_X(M,t) R_X(M,t)\equiv \int R^{(1,1)}_{XY}(M,M_Y,{\bm w}, t) \dd M_Y\dd^n{\bm w},
 \label{eq:gamma_definition}
\end{equation}
from which it follows that
$$
 \gamma_X(M,t) R_X(M,t) = \gamma_Y(M,t) R_Y(M,t).
$$
We can then assume that a fraction $\alpha_X$ are in $XX$ binaries so that
\begin{eqnarray}
 R_X^{(1)}(M,t)&=&(1-\alpha_X-\gamma_X) R_X(M,t)\\ \label{eq:RxRy_definition1}
 R_X^{(2)}(M,M,{\bm w}, t) &=&\frac{1}{2}\alpha_X R_X(M,t){\cal P}_X({\bm w}),
 \label{eq:RxRy_definition2}
\end{eqnarray}
where ${\cal P}_X({\bm w})$ is the PDF of the orbital parameters at the time of formation normalized such that
$$
\int {\cal P}_X({\bm w})\dd^n{\bm w}=1,
$$
and ${\cal P}_{XY}({\bm w})$ is the PDF of the orbital parameters of the hybrid systems. We can then check that
\begin{align}
 R_X(M,t) = R^{(1)}_X(M_, t) & + 2\int R_X^{(2)}(M,M,{\bm w}, t)\dd^n{\bm w} \nonumber \\ & + \int R_{XY}^{(1,1)}(M,M_Y,{\bm w}, t)\dd^n{\bm w}\dd M_Y.
\end{align}
Note also that $R_X^{(2)}$ and $R_{XY}^{(1,1)}$ do not have the same dimensions.

\subsubsection{Evolution of the densities}

We shall now formulate the equations that describe the evolution of the density of binary compact objects in our model. As a first step, let us consider the case with only one species. During the evolution of the binary its orbital parameters are constantly changing due to perturbations or the emission of gravitational waves so that their time evolution can be expressed as
\begin{equation}\label{eq:ee1}
 \frac{\dd {\bm w}}{\dd t} = {\bm f}({\bm w}, M)
\end{equation}
where ${\bm f}({\bm w}, M)$ depends on the physical process at work. A merger occurs when ${\bm w}={\bm w}_{\rm merger}$.

At any given time, single objects $X$ of mass $M$ are formed via two routes: direct formation (i.e. SN collapse) with a rate $R_X^{(1)}$ and from the merger of binary systems. In the latter case the final state has a mass $2M-\Delta M$ where $\Delta M$ is a function of $M$ that corresponds to the energy radiated in gravitational waves. We shall define $S(M,M,t)$ as the rate of mergers of binary systems of masses $(M,M)$ at time $t$ and assume that a fraction $(1-\beta_X)$ of the merger products remains single. Binary systems thus form from newly-born objects $X$ at a rate $R_X^{(2)}(M,{\bm w},t)$, as well as from merger remnants with a fraction $\beta_X$. In addition, their orbital parameters evolve according to Eq.~(\ref{eq:ee1}).

These considerations translate into the following set of equations:
\begin{widetext}
\begin{eqnarray}
 \frac{\dd n_X^{(1)}(M,t)}{\dd t} &=& R_X^{(1)}(M,t) +(1-\beta_X)S\left(M',M',t\right)\label{evo2}\\
 \frac{\dd n_{X}^{(2)}(M,M,{\bm w}, t)}{\dd t} &=& R_X^{(2)}(M,M,{\bm w},t)+\frac{1}{2}\beta_X S\left(M',M',t\right){\cal P}_X({\bm w})-  \frac{\partial }{ \partial {\bm w}}.[{\bm f}\left({\bm w},M\right)n_{X}^{(2)} \left(M,M,{\bm w},t\right)]\label{evo3}\\
  M&=&2M'-\Delta M (M').\label{evo4}
\end{eqnarray}
\end{widetext}
In Eq.~(\ref{evo3}), the last term describes the evolution of the density of systems in the 2-dimensional space of their orbital parameters. If one thinks of ${\bm f}$ as a velocity and $n_{XX}^{(2)}$ as the density of the fluid, then one recognizes a continuity equation with a sink and a source term. Eq. (\ref{evo3}) relates the mass of the two merging stars $M'$ to the mass of the final state $M$. The source term $S$ due to the mergers takes the form
\begin{equation}
 S\left(M',M',t\right)=\int_{C_m}  {\bm f}n^{(2)}_X\left(M',M',{\bm w},t\right).\dd{\bm \ell}
 \label{eq:source_term}
\end{equation}
where $C_m$ is a contour in the 2-dimensional parameter space around ${\bm w}_{\rm merger}$ so that $\dd{\bm \ell}$ has dimension $w$. $C_m$ characterizes all the systems that merge in a time step.\\


We note that a similar approach which utilizes the continuity equation was taken by Refs. \citep{1994MNRAS.268..841B,1995MNRAS.274..115M,2001MNRAS.327..531I} to study the GW background from high-mass binary pulsars in our Galaxy. In particular, these studies assumed a \emph{steady-state} solution, which leads to a particular distribution of orbital parameters. In the present analysis our goal is to calculate the source distribution on cosmological timescales, and we therefore relax the steady-state assumption and introduce the source and sink terms.\\

In order to include hybrid systems, we need to determine the final state of a $XY$ merger. In the following we assume that it leads to the formation of a single object $X$ (that is, a BH) of mass $M'_X=M_X+M_Y-\Delta M$. The evolution of the $XY$ systems is then similar to Eq.~(\ref{evo3}),
\begin{widetext}
\begin{eqnarray}
   \frac{\dd n_{XY}^{(1,1)}(M_X,M_Y,{\bm w}, t)}{\dd t} = R_{XY}^{(1,1)}(M_X,M_Y,{\bm w},t)
   -\frac{\partial }{\partial {\bm w}}.[{\bm f}_{XY}\left({\bm w},M_X,M_Y\right)n_{XY}^{(1,1)} ] \left(M_X,M_Y,{\bm w},t\right),
   \label{eq:dnxy}
\end{eqnarray}
\end{widetext}
where ${\bm f}_{XY}\left({\bm w},M_X,M_Y\right)$ describes the evolution of the orbital parameters of $XY$, that may be different from $XX$ systems, and where we assume for simplicity that second generation $X$ do not form new $XY$ systems. At merger, we get a source term
\begin{equation}
 S_{XY\rightarrow X}\left(M',t\right)=\int_{C_m}  {\bm f}_{XY}n^{(1,1)}_{XY}\left(M_X,M_Y,{\bm w},t\right).\dd{\bm \ell}\dd M_Y
 \label{eq:sxy}
\end{equation}
with the constraint $M'_X=M_X+M_Y-\Delta M (M_X,M_Y)$. This source has to be added to Eq. (\ref{evo2}) which describes the evolution of $n^{(1)}_X$, assuming the products of a $XY$ merger remain single (see Eq. (\ref{eq:evo1_full}) below).

\subsubsection{Evolution of the orbital parameters}

Compact binaries undergo orbit circularization due to emission of GW. The evolution of the eccentricity $e$ and the semi-major axis $a$ is given by Ref. \citep{PhysRev.131.435}
\begin{eqnarray}
 \frac{\dd a}{\dd t} &=& -\frac{64}{5}\frac{G^3\mu m^2}{c^5a^3}\frac{\left(1+\frac{73}{24}e^2 + \frac{37}{96}e^4\right)}{(1-e^2)^{7/2}} \label{eq:orbital_ev}\\ 
 \frac{\dd e}{\dd t} &=& -\frac{304}{15}\frac{G^3\mu m^2}{c^5a^4}\frac{e\left(1+\frac{121}{304}e^2\right)}{(1-e^2)^{5/2}}
\end{eqnarray}
where $m=M_1+M_2$ is the total mass of the binary and $\mu=M_1M_2/M$ is the reduced mass.
Clearly when $e=0$, $\dd e/\dd t=0$ so that a circular orbit remains circular. 

The lifetime of a binary system is given by
\begin{equation}
 \tau(a_0,e_0) = \frac{5}{256}\frac{c^5a_0^4}{G^3m^2\mu}F(e_0)
 \label{eq:lifetime}
\end{equation}
where
\begin{equation}
F(e_0)=\frac{48}{19}\frac{1}{g^4(e_0)}\int_0^{e_0}\frac{g^4(e)(1-e^2)^{5/2}}{e(1+\frac{121}{304}e^2)}\dd e\:
\end{equation}
and
\begin{equation}
g(e)= \frac{e^{12/19}}{1-e^2} \left(1+\frac{121}{304}e^2\right)^{870/2299}.
\end{equation}

\begin{figure}[htbp]
\begin{center}
\includegraphics[width=.9\columnwidth]{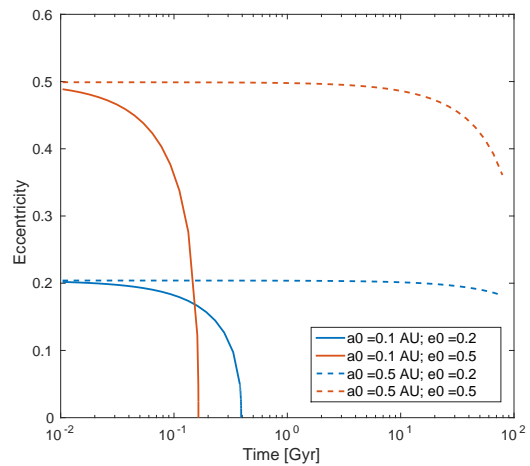}
\caption{Evolution of eccentricity for a $30M_{\odot}-30M_{\odot}$ binary with different initial eccentricities and separations. These different initial conditions induce the difference in merging timescales of several orders of magnitude.}
\label{fig:orbParam}
\end{center}
\end{figure}

A solution of equations (\ref{eq:orbital_ev}) in the case of a $30M_{\odot}-30M_{\odot}$ binary is shown in Fig. \ref{fig:orbParam}. Note the difference in merging timescale: from $0.2$ Gyr for the case $e_0=0.5,a_0=0.1$ AU to more than the age of the Universe for $e_0=0.5$ but $a_0=0.5$ AU. It is clear from these results that the initial distribution of orbital parameters will have a significant influence on the merger rate of binary compact objects. 

Note that the evolution of the orbital parameters can be more complex if the binary is embedded in a dense stellar environment. This is the case for stellar-mass BBH formed in globular clusters \citep{2016arXiv160604889A} and SMBHs at sub-parsec separations \citep{2008ApJ...686..432S}. In both these cases the binary may enter the observable frequency band while still having a non-negligible eccentricity. 

\begin{figure}[htbp]
\begin{center}
\includegraphics[width=.9\columnwidth]{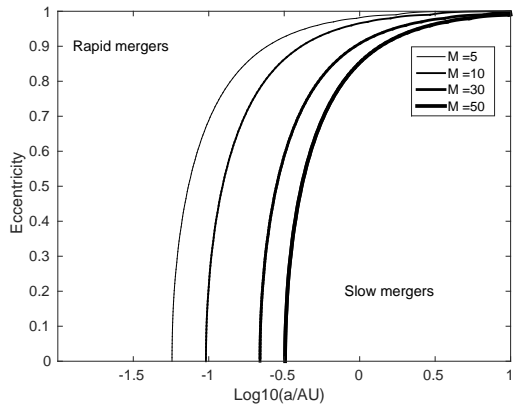}
\caption{The region in parameter space (eccentricity and semi-major axis) corresponding to \emph{rapid mergers} (i.e. within the age of the Universe) to the left of the solid lines and \emph{slow mergers} (i.e. taking longer than the age of the Universe) to the right of the solid lines. Each line corresponds to a different mass of the BBH components, as indicated in the legend.}
\label{fig:merging_time}
\end{center}
\end{figure}

The solution to eq. (\ref{eq:lifetime}) for several masses ($5M_{\odot}-5M_{\odot}$, $10M_{\odot}-10M_{\odot}$, $30M_{\odot}-30M_{\odot}$ and $50M_{\odot}-50M_{\odot}$ binaries) is shown in Fig. (\ref{fig:merging_time}): the area to the left of each curve indicates the parameter space for which the merger occures within the age of the Universe. The merger time is clearly very sensitive to the initial semi-major axis, but also to the masses and, to a lesser extent, the eccentricity.

\subsection{Complete synthetic model of the evolution of binary compact objects}
\label{sec:complete_set}

\begin{figure}
\begin{center}
\includegraphics[width=1.\columnwidth]{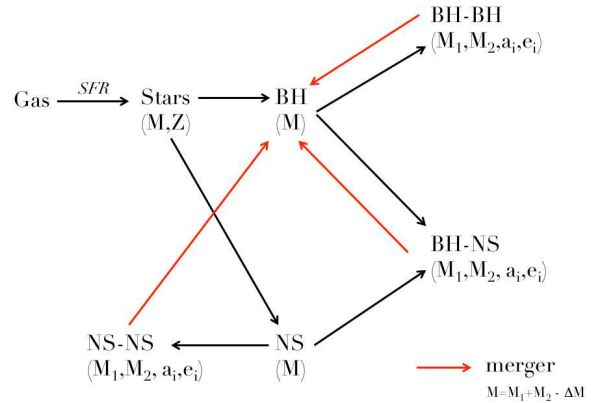}
\caption{A schematic representation of our model framework. Stars form from interstellar gas with metallicity $Z$ as described in an underlying model of galactic evolution. Massive stars end their lifes as BHs or NSs, according to their initial mass and metallicity. We assume that part of these objects form binary systems: $BH-BH$, $NS-NS$ and $BH-NS$. Gravitational waves are produced during the inspiral phase and merger of binary systems prior to their collapse into a BH (marked by red arrows) and during the core collapse of massive stars. The latter also affect the metallicity of the interstellar gas.}
\label{fig-1}
\end{center}
\end{figure}

Our model framework is shown schematically in Figure~\ref{fig-1}. The basis for our calculation is a cosmic evolution model that follows the growth of galaxies, in particular their stellar and gaseous components, including the evolution of gas-phase metallicity $Z$. An underlying stellar evolution model describes the fate of massive stars based on their initial mass and metallicity and predicts whether they would form a BH or a NS at the end of their life. We note that other parameters, such as stellar rotation, are expected to strongly influence the evolution of massive stars and thus the mass of the BH or NS that forms. A certain fraction of these objects belong to binary systems, as shown on Figure~\ref{fig-1}. These systems emit gravitational radiation and may experience interactions with their surroundings, and as a result their orbits can shrink and they merge (as indicated by the red arrows). All the mergers we consider are assumed to lead to a formation of a BH. We can now formulate the set of differential equations that govern the evolution of this system using eqs. (\ref{eq:ee1})-(\ref{eq:sxy}).

We consider all three types of binary objects, namely binary NSs, BHs and BH-NS, where the evolution of the orbital parameters of each type is governed by eq. (\ref{eq:ee1}). The evolution of the number densities of binary NSs and BHs are given by Eqs. (\ref{evo2})-(\ref{evo4}) with the source terms provided by eqs. (\ref{eq:source_term}) and (\ref{eq:sxy}) while the evolution of the hybrid population NS-BH is provided by Eq. (\ref{eq:dnxy}).


The resulting equations (\ref{Fevo1}-\ref{Fevo4}) can be found in the Appendix. Their solution provides the evolution of the number densities of single and binary compact objects under a specific set of astrophysical assumptions. We note that these equations are generic in the sense that the processes driving this evolution are not implicit in the formulation. For example, the evolution of the orbital parameters can be primariliy driven by the emission of GW, as we assume below, but can also be influenced by interactions with the surrounding medium. This influence can be taken into account by an introduction of an appropriate function ${\bm f}({\bm w}, M)$ describing this evolution. Similarly, the formation rates should be obtained from specific astrophysical models.


\section{Computation of the gravitational wave background}
\label{sec:GWback}

\subsection{Composite spectrum}
\label{sec:spectrum}

An inspiraling system of compact binaries in an elliptical Keplerian orbit emits gravitational radiation in a discrete set of harmonics of the orbital frequency $f_n=nf_0$ for $n \geq 1$ where $(2\pi f_0)^2=Gm/a^3$. Note that, as we have seen, $a$ (and hence $f_0$) evolves with time due energy losses to gravitational wave emission. Instanteneously, the power emitted by the inspiraling binary at frequency $f_n$ is \citep{PhysRev.131.435}:
\begin{equation}
 \left. \frac{dE}{dt}\right\vert_n = \frac{32G^4 \mu^2 m^3}{5c^5 a^5}g(n,e)
\end{equation}
where
\begin{equation}
 g(n,e) = \frac{n^6}{96 a^4}\left[A^2_n(e)+B^2_n(e)+3C^2_n(e)-A_n(e)B_n(e) \right]\:.
\end{equation}
The functions $A_n,B_n,C_n$ are given in terms of Bessel functions \citep{GravBookMaggiore}:
\begin{eqnarray}
 A_n(e) = & \frac{a^2}{n}\lbrace J_{n-2}(ne) -J_{n+2}(ne)\nonumber \\ & -2e\left[J_{n-1}(ne)-J_{n+1}(ne) \right] \rbrace
\end{eqnarray}
\begin{equation}
 B_n(e) = \frac{(1-e^2)a^2}{n}\lbrace J_{n+2}(ne) - J_{n-2}(ne) \rbrace
\end{equation}
\begin{eqnarray}
 C_n(e) = & \frac{\sqrt{1-e^2}a^2}{n}\lbrace J_{n+2}(ne) + J_{n-2}(ne) \nonumber \\ &  - e\left[J_{n+1}(ne) - J_{n-1}(ne) \right] \rbrace .
\end{eqnarray}

The total emitted power is then given by the sum over all the harmonics:
\begin{equation}
 \frac{dE}{dt}=\sum_{n=1}^{\infty}\left.\frac{dE}{dt}\right\vert_n\:.
 \label{eq:dedt}
\end{equation}

The energy per unit \emph{emitted} frequency $f$ (we omit the subscript in this section) by a system with orbital parameters $(a,e)$ can be expressed as:
\begin{equation}
 \frac{dE}{df} = \frac{dE}{dt}\left(\frac{df_0}{dt} \right)^{-1}\frac{df_0}{df}\:.
 \label{eq:dedf1}
\end{equation}
The orbital frequency evolves with time as $df_0/dt=-(3f_0/2a)df/dt$ where
$da/dt$ is given by Eq. (\ref{eq:orbital_ev}). Furthermore, $df_0/df=1/n$ for $f=nf_0$ and $0$ otherwise. Combining Eqs. (\ref{eq:dedt}) and (\ref{eq:dedf1}) 
and defining:
\begin{equation}
    \delta(x,y)=
    \begin{cases}
      1, & \text{if}\ x=y \\
      0, & \text{otherwise}
    \end{cases}
  \end{equation}
we obtain the following expression for the energy spectrum:
\begin{equation}
 \frac{dE}{df}=\frac{64G^4\mu^2 m^3}{15c^5 a^4 f_0}\sum_{n=1}^{\infty}\frac{g(n,e)}{n}\left|\left(\frac{da}{dt} \right)^{-1} \right|\delta\left(\frac{f}{f_0},n\right) \:.
\end{equation}
Finally, using Eq. (\ref{eq:orbital_ev}), the usual definition of the chirp mass $M_c=m^{2/5}\mu^{3/5}$ and $(2\pi f_0)^2=Gm/a^3$ we obtain:
\begin{widetext}
\begin{eqnarray}
 \frac{dE}{df}(a,e,M_c) =\frac{(2\pi)^{2/3} (GM_c)^{5/3}}{3G}\frac{(1-e^2)^{7/2}}{\left(1+\frac{73}{24}e^2 + \frac{37}{96}e^4 \right)}\frac{1}{f_0^{1/3}} \sum_{n=1}^{\infty}\frac{g(n,e)}{n}\delta\left(\frac{f}{f_0},n\right)
 \label{eq:dedf_ecc}
\end{eqnarray}
\end{widetext}
where $f$ is the emitted frequency and the notation emphasizes that the spectrum is a function of the \emph{instantaneous} orbital parameters $(a,e)$ and the chirp mass of the binary (Eq. (\ref{eq:dedf_ecc}) is equivalent to Eq. (15) of Ref. \citep{2015MNRAS.449.2700E}). The sum in Eq. (\ref{eq:dedf_ecc}) can be approximated by an analytical expression (Ref. \citep{2015PhRvD..92f3010H} and see also \citep{2007PThPh.117..241E}).
In the case of a circular orbit $g(n,e=0)=\delta(2,n)$ and we recover the familiar expression:
\begin{equation}
 \frac{dE}{df}=\frac{\pi^{2/3}(GM_c)^{5/3}}{3G}\frac{1}{f^{1/3}} \:.
\end{equation}

The energy spectrum is therefore given by a \emph{frequency comb}, which becomes more and more sharply peaked around $n=2$ as the eccentricity decreases. Note also that as the semi-major axis shrinks, the orbital frequency $f_0$ grows and the whole spectrum is shifted to higher frequencies.

Eq. (\ref{eq:dedf_ecc}) describes the inspiraling phase of binaries on eccentric orbits and is valid only up to some frequency $f_{merger}$. Since by this time the orbit had already circularized, we use the expressions from Ref. \citep{2008PhRvD..77j4017A} for the merger and ringdown phases to describe this stage of the binary evolution: 
\begin{equation}
\frac{\textrm{d}E}{\textrm{d}f}=\frac{(G\pi)^{2/3}M_c^{5/3}}{3}
 \begin{cases}
    f^{-1/3},&  f\leq f_1\\
    \omega_1 f^{2/3},  & f_1 < f < f_2\\
    \omega_2 \left(\frac{f}{1+\left(\frac{f-f_2}{\sigma/2} \right)^2} \right)^2, & f_2 \leq f\leq f_3
\end{cases}
\label{eq:dedf_circ}
\end{equation}
The set of
parameters $(f_1,f_2,f_3,\sigma)$, where $f_1,f_2$ correspond to the end of the
inspiral and merger phases, respectively, is taken from Ref.
\citep{2008PhRvD..77j4017A} for the case of non-spinning BHs for each set of
masses. The constants $\omega_1=f_1^{-1}$
and $\omega_2=f_1^{-1}f_2^{-4/3}$ are chosen to make
$\textrm{d} E_{\textrm{gw}}/\textrm{d} f$ continuous.

Eqs. (\ref{eq:dedf_ecc}) and (\ref{eq:dedf_circ}) describe the spectrum of the gravitational wave background as a function of the chirp mass and the orbital parameters for eccentric and circular orbits, respectively. In order to use them in Eq. (\ref{eq:Omega_gw}) we also need to account for the comoving number density of the different sources and their orbital parameter distribution. These quantities need to be computed in the context of an astrophysical model of stellar formation and evolution. Below we describe one particular model, but we stress that our formalism can be applied to a wider class of astrophysical prescriptions.

\subsection{Astrophysical model}
\label{sec:astro}

The birthrate of black holes and neutron stars as a function of mass and redshift (or, equivalently, time) $R_{X}(t,m_{X},{\bm w})$ (in units of events per unit time per unit comoving volume per unit eccentricity per unit semi-major axis) is given by \citep{2016arXiv160404288D}: 
\begin{widetext}
\begin{eqnarray}
 R_{X}(t,m_{X},{\bm w})=\int\psi[t-\tau(m)]\phi(m)\delta(m-g_{X}^{-1}(m_{X})){\cal P}_X({\bm w})\textrm{d}m
 \label{eq:birthrate}
\end{eqnarray}
\end{widetext}
where $\tau(m)$ is the lifetime of a star of mass $m$, $\phi(m)$ is the stellar initial mass function (IMF),
$\psi(t)$ is the cosmic star formation rate (SFR) and $\delta(m)$ is the Dirac delta distribution. ${\cal P}_X({\bm w})$ is the PDF of the orbital parameters at birth. The initial stellar
mass and BH/NS masses are related by the function $m_{X}=g_{X}(m)$ which is implicit in the equation above.
Note that $g_{X}$ also depends on time through its metallicity dependence $Z(t)$. We use the galaxy and stellar evolution models described in Ref. \citep{2016arXiv160404288D} and briefly discuss them below.

We start with a description of the IMF and the SFR. We assume a Salpeter IMF with slope $x=2.35$ in the mass range $0.1-100M_{\odot}$: 
\begin{equation}
 \phi(m) = \frac{dN}{dm} = Am^{-x}
\end{equation}
where $A$ is a normalization constant. We use the functional form of Ref. \citep{2003MNRAS.339..312S} for the SFR:
\begin{equation}
 \psi(z)=\nu\frac{a\exp[b(z-z_m)]}{a-b+b\exp[a(z-z_m)]}
\end{equation}
where $z$ is the redshift. Our fiducial model is a fit to the observations of luminous galaxies compiled by Ref. \citep{2013ApJ...770...57B} and complemented by high-redshift observations from Ref. \citep{2011ApJ...737...90B}. We use the fit parameters given in Ref. \citep{2015MNRAS.447.2575V}, namely $\nu=0.178$ $M_{\odot}$yr$^{-1}$Mpc$^{-3}$, $z_m=2$, $a=2.37$ and $b=1.8$. The evolution of the metallicity of interstellar matter is calculated using the cosmic chemical evolution model described in Refs. \citep{2015MNRAS.447.2575V,2004ApJ...617..693D,2006ApJ...647..773D}. We use metal yields from Ref. \citep{1995ApJS..101..181W} and stellar lifetimes from Ref. \citep{2002A&A...382...28S}. This model reproduces the metallicity evolution of high-redshift damped Ly-$\alpha$ absorbers \citep{2012ApJ...755...89R} as discussed in \citep{2015MNRAS.447.2575V} and is consistent with the optical depth to reionization recently measured by the \emph{Planck} Collaboration \citep{2016arXiv160502985P,2016arXiv160503507P}.

The function $g_{X}(m)$ defines the mass of the BH formed from a star with an initial mass $m$ and in general depends on the metallicity of the star \citep{2008NewAR..52..419V} and its rotation \citep{2009A&A...497..243D,2016A&A...588A..50M,2016MNRAS.458.2634M} which determine the amount of mass loss the star experienced before reaching the core collapse phase. If the star belongs to a binary system its evolution strongly depends on its companion and their possible mass exchange, in particular the common envelope phase. Furthermore, the star, single or in a binary, may belong to a dense star cluster, in which case in can be influenced by dynamical interactions with other stars \citep{2014MNRAS.441.3703Z,2016arXiv160604889A}. In this work we make the simplifying assumption that $g_{X}$ depends only on initial stellar mass and metallicity and use the function computed for models of isolated stellar evolution. We stress, however, that all other dependencies can in principle be expressed by a more general $g_{X}$. In this work we use the models of Ref. \citep{2012ApJ...749...91F} to obtain $g_{X}$, in particular their \emph{delayed} model, as described in Ref. \citep{2016arXiv160404288D}. 

The solution to Eqs. (\ref{eq:ee1})-(\ref{evo3}) depends on the initial distribution of the orbital parameters $\textbf{w}=(a,e)$ of the newborn binary system. Unfortunately, there are no direct observations of this quantity. Using observations of massive (O type) stars (BH/NS progenitors) in Galactic stellar clusters,  \citep{2012Sci...337..444S} deduced a binary fraction of about $0.5$ and estimated the PDFs of orbital periods, mass ratios and eccentricities. In particular, they found a strong preference for small separations, so that the orbital periods were distributed as $P(\log T)\propto (\log T)^{-0.55}$. We stress, however, that these distributions refer to massive stars and not to binary compact objects, for which the initial separations are expected to be much larger. Indeed, it can be seen from Fig. 2 of Ref. \citep{2012Sci...337..444S} that binaries on very close orbits are expected to merge while still on the main sequence, before forming two separate compact objects. For the purposes of this work we adopt the power-law PDFs deduced by Ref. \citep{2012Sci...337..444S} and keep the lower and upper values as free parameters. We model the joint PDF as the product ${\cal P}_X({\bm w})=P(e)P(a)$ and assume the following distribution of initial eccentricities in the range $e\in [0,1]$:
\begin{equation}
 P(e) \propto e^{\kappa}
 \label{eq:Pe}
\end{equation}
with $\kappa=-0.42$ \citep{2015ApJ...814...58D} and orbital periods $T$ (in days) for $\log T\in [\log T_{\textrm{min}},\log T_{\textrm{max}}]$:
\begin{equation}
 P(\log T) \propto (\log T)^{-0.5}\:.
 \label{eq:Pa}
\end{equation}
For the calculations shown below we chose a minimal orbital period corresponding to a semi-major axis of $a_{min}$ in the range $0.1-0.3$ AU and a maximal period corresponding to a semi-major axis of $a_{max}=5000$ AU. Note that while $a_{max}$ has little effect on the results, $a_{min}$ affects the merger rate through eq. (\ref{eq:lifetime}). We chose a range of $a_{min}$ that brackets the uncertainty in the merger rate estimated from Advanced LIGO observations \citep{2016arXiv160203842A}, as can be seen in Figure \ref{fig:gw_bbh_all} and discussed below.

Eq. (\ref{eq:birthrate}) describes the birthrate of neutron stars or black holes under the assumptions outlined above. In order to follow the evolution with time of the number density of \emph{binary} systems we need to account for the fraction of compact objects that form binaries with a given set of orbital parameters and the time evolution of those orbital parameters, given by eqs. (\ref{eq:ee1})-(\ref{evo3}). In the following we assume that all the black holes, as well as all the neutron stars are born in binaries, and that there are no BH-NS binaries. This corresponds to a choice of parameters $\alpha_{BH}=\alpha_{NS}=1$ and $\gamma=0$ in Eqs. (\ref{eq:gamma_definition})-(\ref{eq:RxRy_definition2}). Furthemore, we assume that BHs that formed from a merger of any kind remain single and do not merge with another BH, so that $\beta_{BH}=0$ in eqs. (\ref{evo2})-(\ref{evo3}). We then solve Eqs. (\ref{eq:ee1})-(\ref{evo3}) with the birthrates taken from eq. (\ref{eq:birthrate}) to obtain $n_i(M_{c,i},\textbf{w}_i,z_i)$, the number density of binaries of a given type per unit comoving volume per unit mass per unit $\textbf{w}$, where $\textbf{w}=(a,e)$ signifies the orbital parameters of the binary. The time derivative of this quantity is then used in Eq. (\ref{eq:Omega_gw}) to calculate the gravitational wave background. For simplicity we neglect the energy radiated in GW, setting $\Delta M = 0$ in Eq. (\ref{evo4}). We assume that a binary merges immediately (i.e. in less than one timestep) when the separation is $a<100 R_{*}$ where $R_{*}=2GM/c^2$. For the solution of Eqs. (\ref{eq:ee1})-(\ref{evo3}) we use a grid in the $(e,a)$ space with the corresponding resolution of $\Delta e = 0.025$ and $\Delta \log \tilde{a} = 0.13$ where $\tilde{a}=a/R_{*}$. The timestep was $300$ Myr.

We shall now present the gravitational wave background from inspiraling binary BHs and NSs.

\subsection{Inspiraling and merging binary black holes}
\label{sec:ins_BH}

In this section we will treat two types of sources: inspiraling BBH, i.e. binaries that have not merged during the Hubble time and merging BBH. Their number densities are provided by the solution of Eqs. (\ref{eq:ee1})-(\ref{evo3}) assuming the initial distibution of orbital parameters given by Eqs. (\ref{eq:Pe})-(\ref{eq:Pa}). It can be seen from Eq. (\ref{eq:Pe}) that a large fraction of these binaries are expected to have relatively eccentric orbits, and therefore contribute gravitational radiation in a range of frequencies well above their orbital frequency.

\begin{figure}[htbp]
\begin{center}
\includegraphics[width=.9\columnwidth]{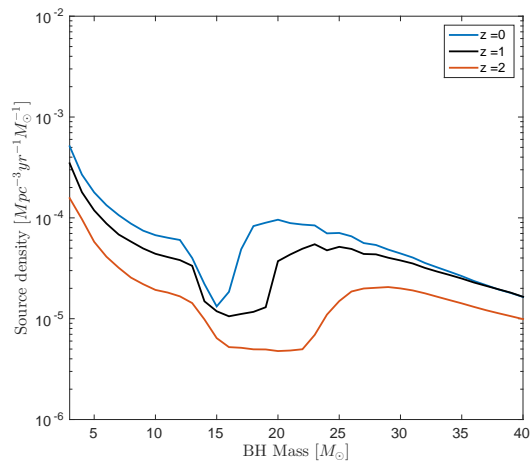}
\caption{The source density ${\dd n_i}/{\dd t}(M,z)$ for our chosen astrophysical model as a function of BH (source frame) mass at $3$ different redshifts. The population builds over time with the number density of BBH with masses around $\sim 20-25M_{\odot}$ growing considerably faster than that of other masses due to transition of stars above $\sim 30M_{\odot}$ from direct collapse, which occures at low metallicities, to SN explosion (and associated lower remnant masses) at higher metallicities.}
\label{fig:astroSourceRates}
\end{center}
\end{figure}

The source density of inspiraling BBH ${\dd n_i}/{\dd t}(M,z)$ (integrated over the orbital paremeters) for our chosen astrophysical model is shown in Figure \ref{fig:astroSourceRates} for $3$ different redshifts. This quantity is the solution of Eqs. (\ref{eq:ee1})-(\ref{evo3}) under the assumptions outlined above and represents the entire population of BBH. The population builds over time, as expected, with the number density of BBH with masses around $\sim 20-25M_{\odot}$ growing considerably faster than that of other masses. The reason for this rapid growth is the evolution in metallicity: whereas at low metallicities stars above $\sim 30M_{\odot}$ undergo direct collapse in the model of Ref. \citep{2012ApJ...749...91F}, these same stars are less efficient in producing massive BHs at higher metallicities and instead end up as BHs with masses around $\sim 20-25M_{\odot}$. This build-up of BHs around $\sim 20-25M_{\odot}$ corresponds to the stagnation in the number density of BHs with $M>30M_{\odot}$.

The energy spectra of inspiraling and merging binaries are given by Eqs. (\ref{eq:dedf_ecc}) and (\ref{eq:dedf_circ}), respectively while Eq. (\ref{eq:Omega_gw}) can be simplified to:
\begin{widetext}
\begin{eqnarray}
\Omega_{\rm gw}(f)&=&\frac{f}{\rho_c c^2 H_0} 
\int  \frac{\dd z}{E(z)}
\int  \dd\textbf{w} \int \dd M_c 
 \frac{\dd n_i}{\dd t}(M_c,\textbf{w},z)
\frac{\dd E^{(i)}_{\rm GW}}{\dd f_e}(M_c,\textbf{w}, f_e)
\end{eqnarray}
\end{widetext}

\begin{figure}[htbp]
\begin{center}
\includegraphics[width=1.\columnwidth]{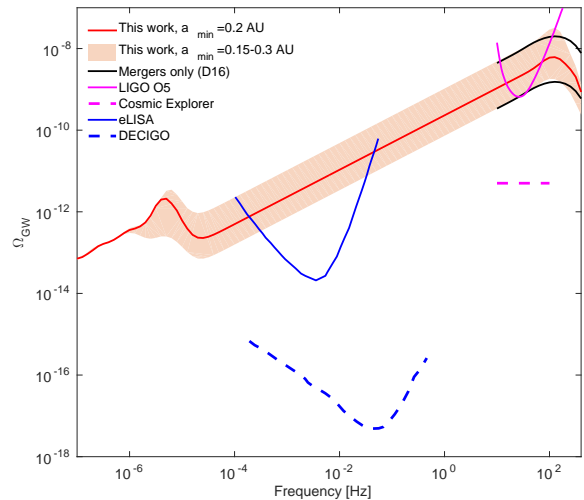}
\caption{GW background from inspiraling and merging binary black holes with $a_{\textrm{min}}=0.2$ AU (red line), the range $a_{\textrm{min}}=0.15-0.3$ AU (red shaded area) and the GW background from merging BBH from Ref. \citep{2016arXiv160404288D} where the merger rate was normalized to the observed value (black lines). Also shown are the expected sensitivities of Advanced LIGO during observing run O5 \citep{2016arXiv160203847T} (solid magenta line), eLISA \citep{2016PhRvL.116w1102S} (solid blue line), DECIGO \citep{2015MNRAS.449.1076P} (dashed blue line) and the Cosmic Explorer \citep{2015PhRvD..91h2001D} (dashed magenta line). The last two curves were estimated from the expected strain sensitivities.}
\label{fig:gw_bbh_all}
\end{center}
\end{figure}

Figure \ref{fig:gw_bbh_all} shows the full energy spectrum resulting from BBHs assuming $a_{\textrm{min}}=0.2$ AU (red line). The low-frequency part of the spectrum is due to \emph{inspiraling} binaries and exhibits a characteristic peak at $f\sim 10^{-5}$ Hz which corresponds to typical separations at which the binary merges within a Hubble time or less. In contrast, high frequencies accessible with Advanced LIGO and VIRGO are dominated by the signal from merging binaries. Unfortunately, the transition frequency is beyond the reach of current and planned GW observatories. We stress, however, that the signal from inspiraling stellar-mass BBH, if detected, will provide important constraints on the population of BBH, complementary to the information that can be obtained from merging systems alone.

The same astrophysical model was used in Ref. \citep{2016arXiv160404288D} to calculate the GW background from merging BBH, where the merger rate was calibrated (uniformly for all masses) to the rate based on the observation of GW150914 \citep{2016arXiv160203842A} of $1.02^{+1.98}_{-0.79} 10^{-7}$ Mpc$^{-3}$yr$^{-1}$ (black lines in Figure \ref{fig:gw_bbh_all}).
The red band in Figure \ref{fig:gw_bbh_all} corresponds to the range $a_{\textrm{min}}=0.15-0.3$ AU and was chosen so as to reflect this uncertainty in the merger rate. In fact, the main uncertainty in our calculation stems from the unknown initial distribution of the orbital parameters which affects the merger rate. Note that below $f\sim 10^{-5}$ Hz this uncertainty does not affect the signal which is due entirely to inspiraling systems.

The amplitude of the peak at the transition frequency $f\sim 10^{-5}$ Hz is determined by the lifetime of the binaries (in other words, the merger rate) and by their distribution of orbital parameters. In particular, since the energy spectrum of an inspiraling binary represents a \emph{frequency comb}, as can be seen from Eq. (\ref{eq:dedf_ecc}), only binaries with the corresponding separations contribute to a given frequency.

We also show in Figure \ref{fig:gw_bbh_all} the expected sensitivities of Advanced LIGO, Cosmic Explorer, eLISA and DECIGO observatories, all of which should be able to detect the stochastic background from merging stellar-mass BBH.

\subsection{Inspiraling binary neutron stars}
\label{sec:ins_NS}

\begin{figure}[htbp]
\begin{center}
\includegraphics[width=.9\columnwidth]{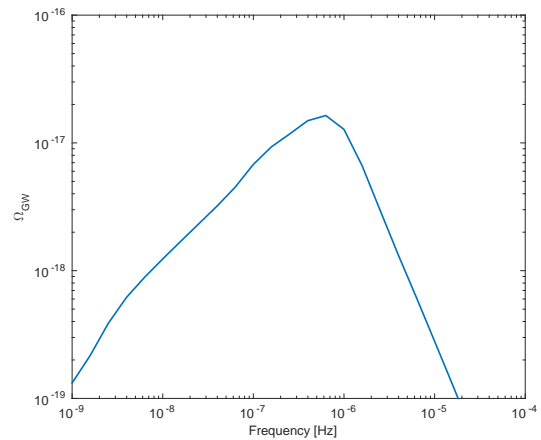}
\caption{Gravitational wave background from inspiraling binary neutron stars that did not merge during the Hubble time.}
\label{fig:gw_bns}
\end{center}
\end{figure}

Similarly to BBH, binary NSs also produce a GW signal during the inspiraling phase. We calculated the contribution from binary NSs in our model, assuming for simplicity a fixed mass for all NSs of $m=2M_{\odot}$. The result, shown in Figure \ref{fig:gw_bns}, is even weaker than in the case of BBH and, as expected, shifted to lower frequency. The signal we compute is much weaker than in Ref. \citep{2015MNRAS.449.2700E}, although note that they introduced a sharp frequency cutoff by hand and also did not use a physically motivated distribution of eccentricities. We do not expect this signal to be detectable with PTA as it is many orders of magnitude below the GW background predicted for merging SMBH.

\section{Discussion}
\label{sec:discussion}

In this paper we developed a synthetic approach that allows to model the evolution of binary compact objects and the GW background they produce. We described the evolution of the number density of binaries and their interactions in the space of orbital parameters with a set of continuity equations providing also for a source and sink terms due to formation and mergers of binary systems. While we used a specific astrophysical model to estimate the GW background from inspiraling and merging BBH, our approach is modular and any ingredient can be updated or tested against competing models. In particular, we can study the rate of formation of compact objects resulting from different astrophysical models, the rate of binary formation, the distribution of initial orbital parameters and the time evolution.

We use our approach to calculate for the first time the GW background from inspiraling BBH. The signal we predict is very weak and, moreover, is not in the frequency range of any of the current or planned GW observatories. Nevertheless, the characteristic shape of the transition from inspiral to merger dominated signal might provide very interesting constraints on the entire population of BBH. Since the majority of BBH are not expected to merge within a Hubble time and are thus beyond the observational capabilites of ground-based interferometers such as Advanced LIGO and VIRGO, the signal we predict offers a unique handle on the properties of this population.

We stress that this work does not include an exhaustive treatment of the various astrophysical effects, such as BH production mechanisms, the effects of rotation, binary co-evolution and possible influence of dense environments. Our treatment of the various GW backgrounds is also far from complete, as we did not include the contribution from SN collapse nor merging NSs. Moreover, we assumed that all the binaries consist of equal-mass objects and did not calculate the background due to inspiraling and merging BH-NS binaries. An extensive study of these topics and the estimate of the associated uncertainties are left for future work.

Finally, we expect this framework to be useful for different classes of GW sources not discussed here, such as SMBH binaries during the later stages of the merger. The signal from inspiraling SMBH which take longer than the age of the Universe to merge falls in the frequency range accesible with PTA ($10^{-9}-10^{-8}$ Hz), while merging SMBH will be observable with eLISA. The evolution of the eccentricity of these systems may be affected by their environment and some of the SMBH binaries may enter the observable frequency band while still retaining a non-negligible eccentricity which will have an imprint on the GW background \citep{2007PThPh.117..241E,2008ApJ...686..432S,2016ApJ...817...70T}. These questions can be treated within the formalism described in this paper and we plan to study them in future work.

\section*{Acknowledgments}                                                      
This work has been done within the Labex ILP (reference ANR-10-LABX-63), part of the Idex SUPER, and received financial state 
aid managed by the Agence Nationale de la Recherche, as part of the programme Investissements d'avenir under the reference ANR-11-IDEX-0004-02.. The work of ID and JS was supported by the ERC Project No. 267117 (DARK) hosted by Universit\'{e} Pierre et Marie Curie (UPMC) - Paris 6, PI J. Silk. JS acknowledges the support of the JHU by NSF grant OIA-1124403. 




\bibliography{fo}

\vspace{5mm}
\appendix
\section{Evolution of number densities of binary compact objects}

The equations describing the evolution of the number densities of binary NSs, BH and BH-NS can be obtained by using eqs. (\ref{eq:ee1})-(\ref{eq:sxy}), as outlined in Section \ref{sec:complete_set} (see also Figure \ref{fig-1}).

We set $X$=BH and $Y$=NS and assume for simplicity that $XX$ and $YY$ binary systems consist of objects of equal masses and that BHs that originate from mergers do not form new binary systems. It follows that the evolution of single and binary BHs and NSs is described by the following set of equations:
\begin{widetext}
\begin{eqnarray}
 \frac{\dd {\bm w}_1}{\dd t} &=& {\bm f}_{XX}({\bm w}_1, M)\label{Fevo1}\\
 \frac{\dd {\bm w}_2}{\dd t} &=& {\bm f}_{YY}({\bm w}_2, M)\label{Fevo1b}\\
 \frac{\dd {\bm w}_3}{\dd t} &=& {\bm f}_{XY}({\bm w}_3, M_X,M_Y)\label{Fevo1c}\\
 \frac{\dd n_X^{(1)}(M,t)}{\dd t} &=& [1-\alpha_X-\gamma_X(M)] R_X(M,t) +(1-\beta_X)S_{XX\rightarrow X}\left(M,M,t\right)\nonumber\\
 &&\qquad+S_{YY\rightarrow X}\left(M,t\right)+S_{XY\rightarrow X}\left(M,t\right)\label{eq:evo1_full}\\
\frac{\dd n_Y^{(1)}(M,t)}{\dd t} &=& [1-\alpha_Y-\gamma_Y(M)] R_Y(M,t) \label{Fevo2}\\ 
 \frac{\dd n_{X}^{(2)}(M,M,{\bm w}_1, t)}{\dd t} &=& \frac{1}{2}\alpha_X R_X(M,t){\cal P}_X({\bm w}_1)+\frac{1}{2}\beta_X S_{XX\rightarrow X}\left(M,M,t\right){\cal P}_X({\bm w}_1)\nonumber\\
   &&\qquad -  \frac{\partial }{ \partial {\bm w}}.[{\bm f}_{XX}\left({\bm w}_1,M\right)n_{X}^{(2)} \left(M,M,{\bm w}_1,t\right)]\label{Fevo3}\\
 \frac{\dd n_{Y}^{(2)}(M,M,{\bm w}_2, t)}{\dd t} &=& \frac{1}{2}\alpha_Y R_Y(M,t){\cal P}_Y({\bm w}_2)\nonumber\\
 &&\qquad-  \frac{\partial }{ \partial {\bm w}_2}.[{\bm f}_{YY}\left({\bm w}_2,M\right)n_{X}^{(2)} \left(M,M,{\bm w}_2,t\right)]\label{Fevo3b}\\
     \frac{\dd n_{XY}^{(1,1)}(M_X,M_Y,{\bm w}, t)}{\dd t} &=& R^{(1,1)}_{XY}(M_X,M_Y,{\bm w}_3,t)\nonumber\\
     &&\qquad-\frac{\partial }{\partial {\bm w}_3}.[{\bm f}_{XY}\left({\bm w}_3,M_X,M_Y\right)n_{X}^{(2)} ] \left(M_X,M_Y,{\bm w}_3,t\right),\\
   S_{XX\rightarrow X}\left(M,M,t\right)&=&\int_{C_m}  {\bm f}_{XX}({\bm w}_1,M')n^{(2)}_X\left(M',M',{\bm w}_1,t\right).\dd{\bm \ell}  \\
   S_{YY\rightarrow X}\left(M,M,t\right)&=&\int_{C_m}  {\bm f}_{YY}({\bm w}_2,M')n^{(2)}_X\left(M',M',{\bm w}_2,t\right).\dd{\bm \ell}  \\
     M&=&2M'-\Delta M (M').\label{Fevo4}\\
   S_{XY\rightarrow X}\left(M_X,t\right)&=&\int_{C_m}  {\bm f}_{XY}n^{(1,1)}_{XY}\left(M'_X,M_Y,{\bm w}_3,t\right).\dd{\bm \ell}\dd M_Y,\\
      M_X&=&M'_X+M_Y-\Delta M (M'_X,M_Y).
\end{eqnarray}
\end{widetext}

This idealized model requires $3$ free parameters: $\alpha_X,\beta_X,\alpha_Y$ that characterize the ratio between the different populations and 3 PDFs: ${\cal P}_{X},{\cal P}_{Y},{\cal P}_{XY}$. These quantities should be provided by an astrophysical model of binary formation.

In addition, the formation rates $R_X$, $R_Y$ and $R_{XY}$ are obtained from the SFR and the stellar evolution model (where $R_X$ and $R_{XY}$ are related through the parameter $\gamma_X$, see eq. (\ref{eq:gamma_definition})).\\

\end{document}